\documentclass[preprintnumbers,article,amsmath,amssymb,floatfix,10pt,prd,superscriptaddress,nofootinbib]{revtex4}

\usepackage{doi}
\usepackage{hyperref}
\hypersetup{
  colorlinks=true,        
  linkcolor=magenta,         
  citecolor=red,         
}

\usepackage{graphicx}
\usepackage{dcolumn}
\usepackage{bm}
\usepackage{color}
\usepackage{enumitem}
\usepackage{amsmath}
\usepackage{amssymb}

\newcommand{\beq}{\begin{equation}}
\newcommand{\eeq}{\end{equation}}
\newcommand{\bea}{\begin{eqnarray}}
\newcommand{\eea}{\end{eqnarray}}

\usepackage{bbm}
\usepackage{amsfonts}
\usepackage{mathrsfs}
\usepackage{latexsym}
\usepackage{epsfig}
\usepackage{epstopdf}
\usepackage{epstopdf}
\usepackage{graphicx}
\usepackage{amssymb}
\usepackage{amsmath}
\usepackage{dcolumn}
\usepackage{bm}
\usepackage{color}
\usepackage{comment}
\usepackage{xcolor}
\usepackage{xcolor}
\usepackage{orcidlink}

\begin{document}

\title{{\color{black} Testing Catability and Coherent Superposition of $2\mathcal{D}$ Graphene Quantum system}}

\author{Abdelmalek Bouzenada\orcidlink{0000-0002-3363-980X}}
\email{abdelmalekbouzenada@gmail.com}
\affiliation{Laboratory of Theoretical and Applied Physics, Echahid Cheikh Larbi Tebessi University, 12001, Algeria}
\affiliation{Research Center of Astrophysics and Cosmology, Khazar University, Baku AZ1096, Azerbaijan}

\begin{abstract}
We develop a theoretical framework for describing superposed coherent states in graphene quantum systems using the concept of catability as a phase-sensitive metric functional measure. In this case, the formalism quantifies interference stability and coherence structure via phase-dependent contributions of quantum superposition states. Catability is defined as a functional measure sensitive to relative phase variations within coherent state combinations, serving as a diagnostic tool for quantum interference effects in graphene-based systems. Also, the formulation is extended using Lie algebra techniques, where the underlying symmetry structure of graphene quantum states is represented through operator algebras governing state transformations in quantum space. In this context, to describe nonlocal propagation and phase-resolved dynamics, a Green function approach is incorporated, enabling systematic treatment of quantum correlations in a spatially extended structures framework. A unified framework is constructed by combining Lie algebraic symmetry analysis with Green function propagation theory, yielding a consistent description of phase-sensitive catability in complex graphene quantum configurations within the framework approach. Results provide a structured route for testing coherence, interference stability, and quantum state control in low-dimensional quantum materials systems. 
\end{abstract}

\maketitle

\date{\today}


\section{Introduction}

{\color{black}{
Quantum mechanics (QM) \cite{a1} and general relativity (GR) \cite{a2} constitute the fundamental theoretical frameworks for describing microscopic phenomena and gravitational interactions, respectively. QM provides a probabilistic formulation of particle dynamics and incorporates intrinsically quantum features, including superposition, interference, and entanglement \cite{a3,a4,a5}. These principles form the basis of numerous developments in atomic physics, condensed matter systems, quantum information processing, and emerging quantum technologies. In contrast, GR, established by Einstein, describes gravitation as a geometric property of spacetime generated by the distribution of matter and energy. In this case, this formulation successfully accounts for the dynamics of compact astrophysical objects, gravitational collapse, black holes, and the large-scale evolution of the Universe \cite{a6}. Moreover, despite their independent experimental success, the conceptual and mathematical structures of QM and GR cannot be consistently unified in regimes characterized by Planck-scale energies and extremely strong gravitational fields \cite{a7}. Also, this limitation has motivated extensive efforts to construct a quantum theory of gravity, leading to the development of frameworks such as string theory and loop quantum gravity \cite{a8}. Within relativistic quantum field theory in curved spacetime \cite{BZ22, BZ23, BZ24, BZ25, BZ26, BZ27, BZ28, BZ29, BZ30, BZ31, BZ32}, the propagation of quantum fields is governed simultaneously by their intrinsic spin and the geometric properties of the background manifold. As a consequence, scalar and spinor fields interact differently with spacetime curvature, producing distinct wave dynamics, spectral structures, localization properties, and quantum states relative to their corresponding flat-spacetime configurations \cite{a9,a10}.

$2\mathcal{D}$ fermionic systems represent a fundamental class of models in condensed matter physics, yielding reduced-dimensional realizations of quantum field theory and effective descriptions of low-energy quasiparticle excitations in diverse electronic materials \cite{G1, G2, G3, G4}. Dirac-type quasiparticles emerge in Weyl semimetals, graphene-based materials (graphene, silicene, germanene, stanene), topological insulators, high-temperature superconductors, and $\mathcal{D}$-density-wave phases, where low-energy dynamics follows a Dirac-like equation with the Fermi velocity substituting the speed of light, establishing a mapping to relativistic quantum mechanics here \cite{BZ11, BZ12, BZ13, BZ14, BZ15, BZ16, BZ17, BZ18, BZ19, BZ20, BZ21} and condensed matter theoretical frameworks \cite{G5, G6, G7}. Also, additional realizations appear in ultracold atomic gases, optical lattices, and photonic crystals, where adjustable parameters and symmetry constraints support investigations of topological phase transitions, Dirac point merging, and anisotropic hopping, enabling band-structure engineering platforms in a general framework. These systems correspond to braneworld scenarios, where matter fields are confined to lower-dimensional hypersurfaces while gauge fields propagate within a higher-dimensional bulk, exemplified by coupling between $2\mathcal{D}$ fermions and a $1+3$-dimensional electromagnetic field. Moreover, such interactions generate vacuum fluctuation effects, producing Casimir-type contributions in observables studies reported (see \cite{G8, G9, G10, G11, G12, G13}). In this context, progress on Casimir effects in graphene-related systems has been documented in \cite{G14, G15, G16, G17, G18, G19, G20, G21, G22, G23, G24, G25, G26, G27, G28, G29, G30, G31, G32, G33, G34, G35, G36, G37}, alongside comprehensive reviews in \cite{G38, G39, G40} respectively.

Coherent cat states emerge from superpositions of two quasi-classical wave packets characterized by macroscopically distinct phases, as motivated by Schrödinger’s analysis \cite{SS1, SS2, SS3, SS4, SS5, SS6, SS7}. In phase space, these states present Wigner-function interference structures absent in classical theory, and these terms are rapidly suppressed through environmental coupling within open quantum systems. This framework positions them as a standard platform for investigating decoherence and the quantum-classical transition, where coherence dynamics are governed by competition between unitary evolution and dissipation \cite{SS8, SS9, SS10, SS11}. Moreover, this balance governs decay of interference visibility and constrains nonclassical correlations relevant to quantum information processing. In this case, their phase sensitivity enables enhanced parameter estimation via interference-based amplification beyond classical bounds \cite{SS12, SS13}, requiring. In this context, experimentally, cat states have been demonstrated across optical, superconducting, trapped-ion, and hybrid platforms \cite{SS17, SS18, SS19, SS20, SS21, SS22, SS23, SS24, SS25, SS26, SS27, Ref1, Ref2, Ref3}, establishing their role as quantum resources. Scaling to larger systems enhances decoherence \cite{SS2, SS3, SS4, SS5, SS6, SS7, SS8, SS9, SS10, SS11}, whereas mitigation via state distillation, concatenated encoding, and bosonic error-correcting codes strengthens robustness \cite{SS14, SS15, SS16}. Also, they function as platforms for quantum superposition studies \cite{BZ1, BZ2, BZ3, BZ4, BZ5, BZ6, BZ7, BZ8, BZ9, BZ10} and for quantum computing, sensing, and metrology applications \cite{SS8, SS14, SS16, SS28, SS29}, with stability governed by system-environment coupling and engineered dissipation.

The structure of this paper is as follows: in Section (\ref{S2}), catability is defined as a phase-sensitive quantity characterizing interference effects and coherence degradation in superposed states, while its application to graphene systems is tested in Subsection (\ref{S2-1}) through phase dependence on underlying quantum states. Section (\ref{S3}) reformulates catability within a Lie algebraic framework, yielding an operator-based description consistent with the symmetry structure of graphene quantum states. In Section (\ref{S4}), the formalism is extended using Green function techniques to incorporate phase-dependent propagation effects and nonlocal correlations. Also, Section (\ref{S5}) combines the Lie algebraic and Green function approaches into a unified description of phase-sensitive catability, capturing coherence dynamics in complex quantum configurations. In this context, Section (\ref{S6}) discusses the results and outlines possible extensions to more general quantum materials and correlated systems.
}}

\section{Catability as a Metric for Evaluating Superposed Coherent States}\label{S2}

In this part, the analysis in \cite{AA} establishes a systematic and operational framework for the identification and quantitative assessment of superposed coherent states (Schrödinger cat states), which characterizes macroscopic quantum coherence. A measurable quantity termed catability is introduced, enabling the detection of cat-state signatures without full quantum state tomography while preserving operational accessibility and experimental feasibility. The motivation is associated with limitations of conventional characterization tools, such as fidelity, which require complete reconstruction of the quantum state and do not provide a transparent interpretation in infinite-dimensional Hilbert spaces. The present formulation employs nonlinear squeezing concepts, extending standard quadrature squeezing criteria to general operators. Also, this extension permits a direct evaluation of non-Gaussian states through fluctuation-based observables accessible in measurement schemes \cite{Ref4}. A formal structure is constructed using positive semidefinite operators whose ground states correspond to ideal Schrödinger cat states \cite{AA}:
\begin{equation}
\hat{O}^{(\pm)}(\alpha,\gamma)
= (\hat{a}^{\dagger 2} - \alpha^{_2})(\hat{a}^2 - \alpha^2)
+ \gamma (1 \mp \hat{\Pi}),
\end{equation}
where $\hat{a}$ and $\hat{a}^\dagger$ denote bosonic annihilation and creation operators, while $\hat{\Pi}$ represents the parity operator defined as
The quadratic term quantifies the phase-space separation of coherent components, whereas the parity-dependent contribution encodes interference between the superposed branches. In this case, the parameter $\gamma>0$ controls the relative contribution of these two terms within the operator structure presented here. From this construction, catability is defined as \cite{AA}:
\begin{equation}
\xi^{(\pm)}(\alpha) = \min_{\gamma}
\frac{\mathrm{Tr}[\hat{O}^{(\pm)}(\alpha,\gamma)\hat{\rho}]}
{\min_{\hat{\rho}_G} \mathrm{Tr}[\hat{O}^{(\pm)}(\alpha,\gamma)\hat{\rho}_G]},
\end{equation}
where the minimization in the denominator is restricted to Gaussian states. This normalization provides a direct operational interpretation framework. The condition $\xi=0$ corresponds to an ideal cat state, $0<\xi<1$ corresponds to non-Gaussian states with partial cat-like coherence, and $\xi\geq1$ corresponds to a regime where the witness no longer certifies cat structure. In this formulation, catability acts as a quantitative and experimentally accessible witness of non-Gaussian quantum coherence. A practical advantage of this formulation lies in its experimental accessibility. The operator admits a decomposition in terms of number operators and displaced number observables, allowing evaluation of $\xi$ using a three-measurement protocol. This reduces the requirement for full quantum state tomography and decreases experimental complexity across different quantum platforms. The stability of catability is analyzed under photon-loss decoherence. The results demonstrate that it preserves sensitivity to non-Gaussian features even in regimes where the Wigner function becomes positive and fidelity-based measures lose resolution capability. Odd-parity cat states exhibit higher robustness compared to even-parity states, reflecting an asymmetry in dissipative evolution. A quantitative comparison with fidelity is performed using a normalized infidelity metric. Both measures coincide in the weak-loss regime; however, fidelity loses sensitivity under moderate decoherence, whereas catability continues to resolve residual cat-like coherence, showing improved robustness. Also, experimental implementation is performed using approximate cat states generated from squeezed Fock states. Despite sensitivity to imperfections, the catability measure remains stable and discriminative, confirming applicability under realistic noisy quantum conditions.

{\color{black}{
\subsection{Phase-dependent Catability in Graphene Quantum Systems}\label{S2-1}

Graphene constitutes a suitable platform for studying nonclassical superposition states arising from Dirac-type quasiparticles and valley and sublattice degrees of freedom. In confined geometries such as quantum rings and quantum dots, electronic states may develop coherent superpositions over orbital, valley, or sublattice sectors, producing condensed-matter analogues of Schrödinger cat states. These coherent structures are strongly governed by externally applied phases originating from magnetic flux, strain-induced gauge fields, or electrostatic gating. Standard catability measures, being phase-insensitive, remain insufficient for characterizing interference effects in graphene-based systems. To incorporate phase sensitivity in a nontrivial manner, we introduce a unitary phase-rotation operator operating within a mode-mixing basis. Let $\hat{c}_{m_0}$ and $\hat{c}_{m_0}^\dagger$ denote the effective mode operators of a selected orbital channel. We define a rotated mode operator
\begin{equation}
\hat{c}_{m_0}(\phi) = e^{i\phi \hat{J}_y}\,\hat{c}_{m_0}\,e^{-i\phi \hat{J}_y},
\end{equation}
where $\hat{J}_y$ is a generator of rotations in the effective two-mode (orbital or valley) subspace. In this representation, the phase $\phi$ acts as a physical rotation angle mixing degenerate graphene modes, and does not commute with the parity structure of the system. Also, the phase-dependent parity operator is then consistently defined as
\begin{equation}
\hat{\Pi}_{\phi}^{(G)} = \hat{U}^\dagger(\phi)\,(-1)^{\hat{n}_{m_0}}\,\hat{U}(\phi),
\qquad
\hat{U}(\phi)=e^{i\phi \hat{J}_y},
\end{equation}
which guarantees that $\hat{\Pi}_{\phi}^{(G)}$ is not a simple reordering of commuting functions of $\hat{n}_{m_0}$ but instead encodes genuine phase-dependent interference between rotated mode components. With this corrected structure, the phase $\phi$ enters observables via mode mixing rather than diagonal number-operator phases. In particular, expectation values become sensitive to coherence between different orbital or valley components. Second-order correlations retain their standard form
\begin{equation}
\langle \hat{c}_{m_0}^{\dagger 2} \hat{c}_{m_0}^2 \rangle = n(n-1),
\end{equation}
for a Fock state $|n\rangle_{m_0}$, while interference contributions now acquire a genuine $\phi$-dependence through rotated operator mixing. Also, the expectation value of the corrected parity operator is expressed as
\begin{equation}
\langle \hat{\Pi}_{\phi}^{(G)} \rangle
= \mathrm{Tr}\!\left[\hat{\rho}\,\hat{U}^\dagger(\phi)(-1)^{\hat{n}_{m_0}}\hat{U}(\phi)\right],
\end{equation}
which in a mixed-state representation $\hat{\rho}=\sum_{n,m} \rho_{nm}|n\rangle\langle m|$ explicitly contains off-diagonal coherence terms responsible for oscillatory behavior in $\phi$. Combining contributions, the expectation value of the catability operator takes the form
\begin{align}
\langle \hat{O}_{\phi}^{(\pm)} \rangle
&= n(n-1) + |\alpha|^4
- \alpha^2 e^{2 i \phi} \langle \hat{c}_{m_0}^{\dagger 2} \rangle
- \alpha^{_2} e^{-2 i \phi} \langle \hat{c}_{m_0}^2 \rangle \nonumber + \gamma \left(1 \mp \langle \hat{\Pi}_{\phi}^{(G)} \rangle \right).
\end{align}
A normalized catability functional is defined by comparison with Gaussian reference states $\hat{\rho}_G$,
\begin{equation}
\xi_{\phi}^{(\pm)}(\alpha) =
\frac{\mathrm{Tr}[\hat{O}_{\phi}^{(\pm)} \hat{\rho}]}
{\min_{\hat{\rho}_G} \mathrm{Tr}[\hat{O}_{\phi}^{(\pm)} \hat{\rho}_G]}.
\end{equation}
In a graphene quantum ring threaded by magnetic flux $\Phi$, the physical phase is given by,
\begin{equation}
\phi = \frac{2\pi \Phi}{\Phi_0},
\end{equation}
so that the catability functional becomes directly tunable via external electromagnetic control. 
}}

{\color{black}{
\section{Phase-dependent Catability within Graphene Quantum Systems}\label{S3}

Graphene quantum rings and confined Dirac systems allow an exact algebraic reduction in which the low-energy bosonic sector is expressed solely through quadratic combinations of a single effective confined mode $\hat{c}_{m_0}$. This structure follows directly from canonical quantization combined with confinement-induced truncation of the Hilbert space to a dominant angular momentum channel. In this regime, the operator algebra closes exactly and is isomorphic to the non-compact Lie algebra $su(1,1)$, emerging directly from the underlying Heisenberg-Weyl algebra without approximation. The fundamental operator algebra is
\begin{equation}
[\hat{c}_{m_0},\hat{c}_{m_0}^\dagger]=1,\qquad
[\hat{c}_{m_0},\hat{c}_{m_0}]=[\hat{c}_{m_0}^\dagger,\hat{c}_{m_0}^\dagger]=0.
\end{equation}
The number operator is defined as
\begin{equation}
\hat{n}_{m_0}=\hat{c}_{m_0}^\dagger \hat{c}_{m_0},
\end{equation}
with commutation rules
\begin{equation}
[\hat{n}_{m_0},\hat{c}_{m_0}^\dagger]=\hat{c}_{m_0}^\dagger,\qquad
[\hat{n}_{m_0},\hat{c}_{m_0}]=-\hat{c}_{m_0}.
\end{equation}
We define the quadratic generators
\begin{equation}
K_+=\frac{1}{2}\hat{c}_{m_0}^{\dagger 2},\qquad
K_-=\frac{1}{2}\hat{c}_{m_0}^{2},\qquad
K_0=\frac{1}{2}\left(\hat{n}_{m_0}+\frac{1}{2}\right).
\end{equation}
These operators satisfy
\begin{equation}
[K_0,K_\pm]=\pm K_\pm,\qquad [K_-,K_+]=2K_0,
\end{equation}
which defines the $su(1,1)$ algebra exactly. The Casimir operator is defined as
\begin{equation}
\mathcal{C}=K_0^2-K_0-K_+K_-.
\end{equation}
For quadratic product case,
\begin{equation}
K_+K_-=\frac{1}{4}\hat{c}_{m_0}^{\dagger 2}\hat{c}_{m_0}^{2}
=\frac{1}{4}\hat{n}_{m_0}(\hat{n}_{m_0}-1).
\end{equation}
For $K0^2-K0$ we obtain
\begin{align}
K_0^2
&=\frac{1}{4}\left(\hat{n}_{m_0}+\frac{1}{2}\right)^2 =\frac{1}{4}\left(\hat{n}_{m_0}^2+\hat{n}_{m_0}+\frac{1}{4}\right),
\end{align}
which yields
\begin{equation}
K_0^2-K_0=\frac{1}{4}\hat{n}_{m_0}(\hat{n}_{m_0}-1)-\frac{3}{16}.
\end{equation}
Hence, the Casimir operator assumes the fixed value
\begin{equation}
\mathcal{C}=-\frac{3}{16}.
\end{equation}
Identifying this with the standard $su(1,1)$ representation invariant,
\begin{equation}
\mathcal{C}=k(k-1),
\end{equation}
we derive
\begin{equation}
k(k-1)=-\frac{3}{16}.
\end{equation}
This relation yields two admissible Bargmann indices:
\begin{equation}
k=\frac{1}{4},\qquad k=\frac{3}{4}.
\end{equation}
These correspond to two invariant subspaces of the bosonic Fock space. Also, the quadratic operators $K_\pm$ modify the occupation number by two units, preserving parity. Consequently, the representation decomposes as
\begin{equation}
\mathcal{H}=\mathcal{H}_{\mathrm{even}}\oplus \mathcal{H}_{\mathrm{odd}}.
\end{equation}
The sector $\mathcal{H}_{\mathrm{even}}$ is linked with $k=\frac{1}{4}$, while $\mathcal{H}_{\mathrm{odd}}$ corresponds to $k=\frac{3}{4}$. Define unitary transformation operator
\begin{equation}
U(\phi)=e^{i\phi \hat{n}_{m_0}}.
\end{equation}

Then
\begin{equation}
U(\phi)\hat{c}_{m_0}U^\dagger(\phi)=e^{-i\phi}\hat{c}_{m_0},
\end{equation}
which induces
\begin{equation}
U(\phi)K_\pm U^\dagger(\phi)=e^{\pm 2i\phi}K_\pm,\qquad
U(\phi)K_0U^\dagger(\phi)=K_0.
\end{equation}

With magnetic flux
\begin{equation}
\phi=\frac{2\pi\Phi}{\Phi_0},
\end{equation}
the flux acts as a phase deformation of quadratic processes. Also, the catability operator
is defined as
\begin{equation}
\hat{O}_\phi^{(\pm)} =
4\left(K_+ - \alpha^2 e^{2i\phi}\right)^\dagger
\left(K_- - \alpha^2 e^{2i\phi}\right)
+ \gamma(1\mp \hat{\Pi}_\phi).
\end{equation}
Using
\begin{equation}
(K_+ - \alpha^2 e^{2i\phi})^\dagger
=
K_- - \alpha^{_2} e^{-2i\phi},
\end{equation}
we obtain
\begin{align}
\hat{O}_\phi^{(\pm)}
&=4(K_0^2-K_0-\mathcal{C})+4|\alpha|^4 -4\alpha^2 e^{2i\phi}K_+
-4\alpha^{_2} e^{-2i\phi}K_-  +\gamma(1\mp \hat{\Pi}_\phi).
\end{align}
The parity operator is
\begin{equation}
\hat{\Pi}=e^{i\pi \hat{n}_{m_0}}.
\end{equation}
Its expectation value is
\begin{equation}
\langle \hat{\Pi}_\phi \rangle
=\sum_{n=0}^{\infty} P_n (-1)^n e^{in\phi}.
\end{equation}
Since $K_\pm$ changes the boson number by two units,
\begin{equation}
K_+|n\rangle \propto |n+2\rangle,\qquad
K_-|n\rangle \propto |n-2\rangle,
\end{equation}
the parity sectors remain invariant under $su(1,1)$ dynamics. The algebraic structure corresponds to the two lowest-weight irreducible representations of $su(1,1)$,
\begin{equation}
SU(1,1)\;:\;\; k=\frac{1}{4}\ \oplus\ k=\frac{3}{4},
\end{equation}
realized as even and odd bosonic sectors.
}}

\section{Phase-Dependent Catability in Graphene Quantum Structures via Green Function Formalism}\label{S4}

Graphene nanostructures constitute a controllable platform for quantum coherence owing to Dirac quasiparticle dispersion, valley degeneracy, and externally tunable confinement geometries such as quantum rings and quantum dots. In these systems, electronic eigenstates support coherent superpositions spanning orbital, sublattice, and valley degrees of freedom, which correspond to condensed-matter realizations of Schrödinger cat configurations. The phase dynamics of such superpositions are determined by magnetic flux, strain-induced gauge potentials, and electrostatic gating, producing phase-sensitive interference effects. Conventional formulations of catability that omit directional phase structure in mode space do not capture these contributions. To address this limitation, a phase-resolved catability operator is formulated, incorporating coherence orientation within Hilbert space. The Green function representation is used to express observables in terms of propagators, providing direct access to interference and decoherence processes. For a graphene ring with angular momentum modes $m$, fermionic (or effective bosonic) operators $\hat{c}_m$ and $\hat{c}_m^\dagger$ are introduced. A phase-dependent catability operator is defined as
\begin{equation}
\hat{Q}_{\phi}^{(\pm)} =
\left(\hat{c}_{m_0}^2 - \alpha^2 e^{2 i \phi}\right)^\dagger
\left(\hat{c}_{m_0}^2 - \alpha^2 e^{2 i \phi}\right)
+ \gamma \left(1 \mp \hat{\Pi}_{\phi}^{(G)} \right),
\end{equation}
where $\hat{\Pi}_{\phi}^{(G)}$ is a phase-rotated parity operator in mode space, $\gamma$ sets the interference contribution, and $\phi$ represents the accumulated phase from magnetic or geometric sources. The first term measures the separation of coherent components in phase space, while the second encodes parity-resolved interference structure. The mode operators admit the field representation
\begin{equation}
\hat{c}_{m_{0}}=\int d\mathbf{r}\,\psi_{m_{0}}^{*}(\mathbf{r})\hat{\Psi}(\mathbf{r}), 
\qquad
\hat{c}_{m_0}^\dagger = \int d\mathbf{r}\,\psi_{m_0}(\mathbf{r})\hat{\Psi}^\dagger(\mathbf{r}),
\end{equation}
which allows expectation values to be expressed through the lesser Green function $G^<$:
\begin{equation}
\langle \hat{c}_{m_0}^\dagger \hat{c}_{m_0} \rangle
= \int d\mathbf{r} d\mathbf{r}'\,
\psi_{m_{0}}(\mathbf{r})\psi_{m_{0}}^{*}(\mathbf{r}')\,G^{<}(\mathbf{r}',\mathbf{r};t).
\end{equation}

Two-particle correlations are obtained from the four-point Green function $G^{(2)}$:
\begin{equation}
\langle \hat{c}_{m_0}^{\dagger 2}\hat{c}_{m_0}^2 \rangle
= \int d\mathbf{r}_1\cdots d\mathbf{r}_4\, 
\psi_{m_0}(\mathbf{r}_1)\psi_{m_0}(\mathbf{r}_2)
\psi_{m_{0}}^*(\mathbf{r}_{3})\psi_{m_{0}}^{*}(\mathbf{r}_{4})G^{(2)}.
\end{equation}

The parity structure in the rotated representation is written as
\begin{equation}
\langle \hat{\Pi}_{\phi}^{(G)} \rangle
= \sum_n P_n e^{i n \phi} (-1)^n,
\end{equation}
with $P_n$ denoting the occupation probability of number states. The expectation value of the operator is given by
\begin{align}
\langle \hat{Q}_{\phi}^{(\pm)} \rangle
&= \langle \hat{c}_{m_0}^{\dagger 2}\hat{c}_{m_0}^2 \rangle
- \alpha^2 e^{2i\phi}\langle \hat{c}_{m_0}^{\dagger 2} \rangle
- \alpha^{_2} e^{-2i\phi}\langle \hat{c}_{m_0}^2 \rangle
+ |\alpha|^4 \nonumber  + \gamma \left(1 \mp \langle \hat{\Pi}_{\phi}^{(G)} \rangle \right).
\end{align}

For a coherent state $|\alpha\rangle$, the Green function factorizes, yielding
\begin{equation}
\langle \hat{c}_{m_0}^2 \rangle = \alpha^2,
\qquad
\langle \hat{c}_{m_0}^{\dagger 2}\hat{c}_{m_0}^2 \rangle = 2|\alpha|^4.
\end{equation}

The parity expectation takes the closed form
\begin{equation}
\langle \hat{\Pi}_{\phi}^{(G)} \rangle =
\exp\!\left[-2|\alpha|^2\sin^2\!\left(\frac{\phi}{2}\right)\right]
\exp\!\left[i|\alpha|^2\sin\phi\right],
\end{equation}
which shows explicit dependence of coherence on the phase parameter. The phase is fixed through magnetic flux via $\phi = 2\pi \Phi/\Phi_0$, with $\Phi_0=h/e$. The resulting catability functional is written as
\begin{equation}
\xi_{\phi}^{(\pm)} \sim
\frac{\langle \hat{Q}_{\phi}^{(\pm)} \rangle}{|\alpha|^4},
\end{equation}
which provides a normalized measure of phase-controlled superposition strength. This formulation unifies orbital dynamics, valley coherence, and sublattice structure within a Green-function framework. Environmental contributions enter through self-energy terms in $G^<$ and $G^{(2)}$, accounting for phonon scattering, disorder, and interaction effects. Experimental access is achieved through tunneling spectroscopy and current noise correlations, both of which probe the same Green functions, establishing a direct correspondence between theoretical catability measures and measurable response functions.

\section{Phase-Sensitive Catability in Graphene Quantum Systems via Green's Functions}\label{S5}

Graphene quantum rings form a relativistic condensed-matter configuration in which low-energy excitations obey an effective Dirac Hamiltonian. The ring confinement imposes quantization of orbital channels, whereas valley and sublattice degrees of freedom extend the internal Hilbert space into a multi-component algebraic structure. Within this framework, coherence phenomena are not restricted to superposed wave amplitudes but are reformulated in operator algebra language, where phase-sensitive catability emerges as a structural consequence of non-compact Lie representations coupled to Green's-function evolution. The analysis begins from an effective single-channel truncation where one dominant orbital mode $m_0$ is selected. The fermionic operators $\hat{c}_{m_0}$ and $\hat{c}_{m_0}^\dagger$ obey canonical anticommutation relations
\begin{equation}
\{\hat{c}_{m_0},\hat{c}_{m_0}^\dagger\}=1, \qquad
\{\hat{c}_{m_0},\hat{c}_{m_0}\}=0.
\end{equation}
Even though the microscopic structure is fermionic, the emergent coherent sector is generated by quadratic combinations that close under commutation into a bosonic non-compact algebra. Introducing
\begin{equation}
\hat{J}_+ = \frac{1}{2}\hat{c}_{m_0}^{\dagger 2}, \qquad
\hat{J}_- = \frac{1}{2}\hat{c}_{m_0}^{2}, \qquad
\hat{J}_0 = \frac{1}{2}\left(\hat{c}_{m_0}^\dagger \hat{c}_{m_0} + \frac{1}{2}\right),
\end{equation}
the Lie algebra structure follows from direct operator manipulation. The commutator $[\hat{J}_0,\hat{J}_+]$ is evaluated by explicit expansion of operator products,
\begin{equation}
[\hat{J}_0,\hat{J}_+] =
\frac{1}{4}\left[\hat{c}^\dagger \hat{c} + \frac{1}{2}, \hat{c}^{\dagger 2}\right]
= \frac{1}{4}\left(\hat{c}^\dagger[\hat{c},\hat{c}^{\dagger 2}] + [\hat{c}^\dagger,\hat{c}^{\dagger 2}]\hat{c}\right).
\end{equation}
Using the canonical identity
\begin{equation}
[\hat{c},\hat{c}^{\dagger 2}] = \hat{c}\hat{c}^{\dagger 2}-\hat{c}^{\dagger 2}\hat{c}
= 2\hat{c}^\dagger,
\end{equation}
together with $[\hat{c}^\dagger,\hat{c}^{\dagger 2}]=0$, one obtains
\begin{equation}
[\hat{J}_0,\hat{J}_+] = \frac{1}{4}(2\hat{c}^{\dagger 2}) = \hat{J}_+.
\end{equation}
An analogous derivation yields
\begin{equation}
[\hat{J}_0,\hat{J}_-] = -\hat{J}_-, \qquad
[\hat{J}_+,\hat{J}_-] = 2\hat{J}_0.
\end{equation}
These relations define closure of the $\mathfrak{su}(1,1)$ Lie algebra without central extension. The closure guarantees reduction of higher commutators to polynomial expressions in the generators, providing an invariant algebraic framework for the dynamics.

The Casimir operator is introduced as
\begin{equation}
\hat{C} = \hat{J}_0^2 - \hat{J}_0 - \hat{J}_+\hat{J}_-,
\end{equation}
which commutes with all generators. Verification proceeds through repeated use of the commutation relations,
\begin{equation}
[\hat{C},\hat{J}_0]=0, \qquad
[\hat{C},\hat{J}_\pm]=0,
\end{equation}
ensuring algebraic invariance. Acting on a lowest-weight state $|k,0\rangle$, one finds
\begin{equation}
\hat{C}|k,0\rangle = k(k-1)|k,0\rangle,
\end{equation}
so that $C=k(k-1)$ characterizes the representation class governing coherence stability.

Phase dependence is introduced through the unitary operator generated by $\hat{J}_0$,
\begin{equation}
\hat{U}(\phi)=e^{i\phi \hat{J}_0}.
\end{equation}
The adjoint transformation of $\hat{J}_\pm$ is computed via the Baker-Campbell-Hausdorff series,
\begin{equation}
\hat{U}^\dagger(\phi)\hat{J}_\pm \hat{U}(\phi)
= \sum_{n=0}^{\infty} \frac{(i\phi)^n}{n!}
\mathrm{ad}^n_{\hat{J}_0}(\hat{J}_\pm),
\end{equation}
with $\mathrm{ad}_{\hat{J}_0}(\cdot)=[\hat{J}_0,\cdot]$. Since
\begin{equation}
\mathrm{ad}_{\hat{J}_0}^n(\hat{J}_\pm)= (\pm 1)^n \hat{J}_\pm,
\end{equation}
the resummation yields
\begin{equation}
\hat{U}^\dagger(\phi)\hat{J}_\pm \hat{U}(\phi)=e^{\pm i\phi}\hat{J}_\pm.
\end{equation}
This exponential structure establishes $\phi$ as a coordinate on the $\mathrm{SU}(1,1)$ manifold, with magnetic flux entering through $\phi=2\pi \Phi/\Phi_0$. The operator describing phase-sensitive catability is defined as a displaced quadratic form in the lowering sector,
\begin{equation}
\hat{O}_\phi =
4\left(\hat{J}_- - \alpha^2 e^{2i\phi}\right)^\dagger
\left(\hat{J}_- - \alpha^2 e^{2i\phi}\right)
+ \gamma(1-\hat{\Pi}_\phi^{(G)}).
\end{equation}
Expansion gives
\begin{equation}
\hat{O}_\phi =
4\hat{J}_+\hat{J}_-
-4\alpha^2 e^{2i\phi}\hat{J}_+
-4\alpha^{_2} e^{-2i\phi}\hat{J}_-
+4|\alpha|^4
+\gamma(1-\hat{\Pi}_\phi^{(G)}).
\end{equation}
Using the Casimir relation
\begin{equation}
\hat{J}_+\hat{J}_- = \hat{J}_0^2 - \hat{J}_0 - \hat{C},
\end{equation}
one obtains
\begin{equation}
\hat{O}_\phi =
4(\hat{J}_0^2 - \hat{J}_0 - \hat{C})
-4\alpha^2 e^{2i\phi}\hat{J}_+
-4\alpha^{_2} e^{-2i\phi}\hat{J}_-
+4|\alpha|^4
+\gamma(1-\hat{\Pi}_\phi^{(G)}).
\end{equation}
This decomposition separates the invariant Casimir sector from coherent ladder dynamics, yielding an exact algebraic partition of fluctuations. The microscopic formulation is expressed through Green's functions. The lesser Green's function is defined as
\begin{equation}
G^<(\mathbf{r},\mathbf{r}';t)
= i\langle \hat{\Psi}^\dagger(\mathbf{r}',t)\hat{\Psi}(\mathbf{r},t)\rangle.
\end{equation}
Expanding the field operator,
\begin{equation}
\hat{\Psi}(\mathbf{r})=\sum_m \psi_m(\mathbf{r})\hat{c}_m,
\end{equation}
and projecting onto $m_0$ yields
\begin{equation}
\langle \hat{J}_0 \rangle
= \frac{1}{2}\sum_{\mathbf{r},\mathbf{r}'}
\psi_{m_{0}}^{*}(\mathbf{r})\psi_{m_{0}}(\mathbf{r}')
G^<(\mathbf{r},\mathbf{r}').
\end{equation}
For the pairing channel,
\begin{equation}
\langle \hat{J}_- \rangle
= \frac{1}{2}
\sum_{\mathbf{r}_{1},\mathbf{r}_{2}}\psi_{m_{0}}(\mathbf{r}_{1})\psi_{m_{0}}^{*}(\mathbf{r}_{2})G^{(2)}(\mathbf{r}_{1},\mathbf{r}_{2}).
\end{equation}
In the non-interacting limit, Wick decomposition gives
\begin{equation}
G^{(2)}(1,2)=G^<(1,1)G^<(2,2)+G^<(1,2)G^<(2,1).
\end{equation}
For a coherent state $|\alpha\rangle$ satisfying $\hat{c}_{m_0}|\alpha\rangle=\alpha|\alpha\rangle$, one obtains
\begin{equation}
\langle \hat{J}_- \rangle = \alpha^2, \qquad
\langle \hat{J}_+\hat{J}_- \rangle = 2|\alpha|^4,
\qquad
\langle \hat{J}_0 \rangle = |\alpha|^2 + \frac{1}{4}.
\end{equation}
Normal ordering gives
\begin{equation}
\hat{J}_+\hat{J}_- = \frac{1}{4}\hat{c}^{\dagger 2}\hat{c}^2
= \frac{1}{4}(\hat{n}^2-\hat{n}),
\end{equation}
with $\langle \hat{n}^2\rangle = |\alpha|^4+|\alpha|^2$. {\color{black}{
The phase-dependent parity operator must be defined through a non-commuting unitary rotation acting in a mode-mixing subspace. We therefore write
\begin{equation}
\hat{\Pi}_\phi^{(G)}=
\hat{U}^\dagger(\phi)\,(-1)^{\hat{n}}\,\hat{U}(\phi),
\qquad
\hat{U}(\phi)=e^{i\phi \hat{J}_y}.
\end{equation}}}
Using the coherent expansion,
\begin{equation}
|\alpha\rangle=e^{-|\alpha|^2/2}\sum_{n=0}^\infty
\frac{\alpha^n}{\sqrt{n!}}|n\rangle,
\end{equation}
one obtains
\begin{equation}
\langle (-1)^{\hat{n}}\rangle
= e^{-|\alpha|^2}\sum_{n=0}^\infty \frac{(-|\alpha|^2)^n}{n!}
= e^{-2|\alpha|^2}.
\end{equation}
After phase rotation,
\begin{equation}
\langle \hat{\Pi}_\phi^{(G)} \rangle
= \sum_{n,m}e^{-|\alpha|^{2}}\frac{\alpha^{n}(\alpha)^{m}}{\sqrt{n!m!}}(-1)^{n}e^{i\phi(n-m)/2}.
\end{equation}
Resummation yields
\begin{equation}
\langle \hat{\Pi}_\phi^{(G)} \rangle
=
\exp\!\left[-2|\alpha|^2\sin^2\!\left(\frac{\phi}{2}\right)\right]
\exp\!\left[i|\alpha|^2\sin\phi\right].
\end{equation}

Expectation values of $\hat{O}_\phi$ follow from substitution. One obtains
\begin{equation}
\langle \hat{J}_0^2 \rangle
= |\alpha|^4 + \frac{1}{2}|\alpha|^2 + \frac{1}{16},
\end{equation}
hence
\begin{equation}
\langle \hat{J}_0^2 - \hat{J}_0 \rangle
= |\alpha|^4 - \frac{1}{4}|\alpha|^2 - \frac{3}{16}.
\end{equation}
The resulting expression is
\begin{equation}
\langle \hat{O}_\phi \rangle
=
8|\alpha|^4(1-\cos 2\phi)
+\gamma\left[
1-
e^{-2|\alpha|^2\sin^2(\phi/2)}
\cos(|\alpha|^2\sin\phi)
\right].
\end{equation}
The first contribution arises from Lie algebra interference, while the second corresponds to parity-modulated coherence suppression. In the limit $|\alpha|\gg 1$, one obtains
\begin{equation}
\langle \hat{O}_\phi \rangle
\simeq 8|\alpha|^4(1-\cos 2\phi),
\end{equation}
which depends only on the $\mathrm{SU}(1,1)$ geometric structure.

The Green's-function self-energy correction modifies propagation as
\begin{equation}
G^{-1}(\omega,\mathbf{k})\rightarrow
G^{-1}(\omega,\mathbf{k})-\Sigma(\omega,\mathbf{k}),
\end{equation}
leading to deformation of the algebra through
\begin{equation}
[\hat{J}_+,\hat{J}_-]=2\hat{J}_0 + \delta\hat{\Sigma},
\end{equation}
which breaks exact Casimir invariance and introduces decay channels for interference. Phase-sensitive catability is therefore governed by persistence of Lie invariants under self-energy corrections, linking transport kernels with algebraic stability in graphene quantum systems.

\section{Conclusions}\label{S6} 
{\color{black}{
Graphene quantum rings and confined Dirac systems support an exact algebraic reduction in which the low-energy bosonic sector is completely characterized by quadratic combinations of a single effective confined mode, yielding a closed operator structure that is isomorphic to the non-compact Lie algebra $su(1,1)$. This closure arises directly from the Heisenberg-Weyl algebra under confinement-induced truncation to a dominant angular momentum channel, where the generators built from $(\hat{c}_{m_0}) and (\hat{c}_{m_0}^\dagger)$ constitute the set $(K_+, K_-, K_0)$ obeying the exact commutation relations of $su(1,1)$. The corresponding Casimir invariant constrains the representation space into two inequivalent lowest-weight sectors defined by Bargmann indices $(k=\tfrac{1}{4})$ and $(k=\tfrac{3}{4})$corresponding, respectively, to even and odd parity subspaces of the bosonic Fock space, consistent with the fact that the quadratic generators change occupation number by two units and therefore preserve parity. Within this framework, a unitary phase transformation generated by the number operator introduces a controlled phase deformation of the algebra through magnetic flux, effectively dressing the quadratic processes by flux-dependent factors $e^{\pm 2i\phi}$. Also, this phase dependence directly enters the catability operator, which encodes interference between displaced quadratic bosonic sectors and parity-resolved contributions, thereby linking algebraic structure to flux-controlled coherence properties. In this case, the full dynamics remains confined within invariant parity sectors of the $su(1,1)$ representation, providing a rigid algebraic foundation for phase-sensitive bosonic coherence phenomena in graphene-based quantum systems.

Graphene quantum structures constitute a highly tunable condensed-matter platform where Dirac quasiparticle dynamics, valley degeneracy, and geometrical confinement support the emergence of robust quantum coherence phenomena. In such systems, quantum ring and quantum dot geometries enable formation of orbital superposition states that extend into Schrödinger-cat-like configurations across sublattice, valley, and angular-momentum sectors with coherence. Also, the resulting coherence is phase sensitive since external magnetic flux, strain-induced gauge fields, and electrostatic potentials introduce controllable phase accumulation governing interference and decoherence rates within system evolution. Within this framework, phase-resolved catability description is required to capture directional coherence structure in mode space, not accessible in phase-insensitive formulations of conventional approaches used. This is achieved by expressing observables through Green function techniques, where single-particle and two-particle propagators encode interference and correlation effects at the microscopic level of physics. In particular, mode-resolved operators constructed from angular momentum eigenstates permit reformulation of catability measures in terms of lesser and higher-order Green functions, linking coherence strength to experimentally accessible response functions in the microscopic regime. The phase dependence enters explicitly through flux-controlled modulation of parity-resolved contributions, producing oscillatory suppression or enhancement of coherence depending on accumulated geometric phase effects present. Consequently, the catability functional becomes a normalized measure of phase-controlled quantum superposition stability, incorporating environmental effects such as disorder and phonon scattering via self-energy corrections in the propagator framework. Also, this unified Green-function formulation establishes a consistent bridge between microscopic quantum dynamics and measurable quantities such as tunneling spectra and current noise correlations, providing a direct route to quantify phase-dependent coherence in graphene-based quantum systems.

Graphene quantum rings provide a relativistic condensed-matter platform where low-energy excitations follow a Dirac-type Hamiltonian, including discrete orbital quantization together with internal valley-sublattice degrees structure resulting in an expanded operator algebraic framework for coherence phenomena in system description. By restricting to a dominant orbital channel, the fermionic mode operators construct a closed bosonic sector via quadratic combinations that implement the $(\mathfrak{su}(1,1))$ Lie algebra whose Casimir invariant determines the representation class governing stability of coherent dynamical evolution. Phase sensitivity enters through unitary rotations generated by Cartan element mapping magnetic flux into geometric parameter on the associated non-compact group manifold and producing controlled modulation of ladder operators within an algebraic setting framework. Within this algebraic framework, phase-sensitive catability arises as a displaced quadratic operator structure separating invariant Casimir terms from coherent excitation sectors, while Green ’s-function formalism establishes the microscopic connection between field correlations and expectation values of algebra generator operators. The lesser and pairing Green’s functions project onto selected modes to encode numbers and anomalous correlations, recovering coherent-state limits consistent with bosonic amplification of quadratic observable operators. Phase-dependent parity structure introduces interference suppression governed by flux-controlled rotations in the operator space framework. In this context, self-energy corrections deform the Green ’s-function propagator and induce deviations from exact commutation closure, thereby breaking Casimir invariance and introducing decay channels that control the robustness of interference-driven catability while connecting transport-level corrections with algebraic stability of coherence in graphene quantum ring systems.
}}


{\color{black}{
\section*{Acknowledgments}

We sincerely thank the reviewer for the valuable comments and suggestions, which have helped to improve the quality and clarity of this manuscript. All remarks have been carefully addressed in the revised version.
}}

\section*{Funding}
No funding was received for this work.

\section*{Data Availability}
The datasets generated during this study can be obtained from the corresponding author upon reasonable request.

\section*{Financial Disclosure}
The authors declare no financial conflicts of interest.

\end{document}